\documentclass[11pt,a4paper]{article}

\usepackage[T1]{fontenc} 
\usepackage{amsmath}
\usepackage{amsthm}
\usepackage{amsfonts}
\usepackage{amssymb}
\usepackage{graphicx,psfrag}
\usepackage{hyperref}
\usepackage[font=footnotesize,labelfont=bf]{caption}
\usepackage{amscd}

\usepackage{enumerate}

\begin{document}
\def\thefootnote{\fnsymbol{footnote}}

\begin{center}
\Large{\textbf{A four-dimensional $\Lambda$CDM-type cosmological model\\induced from higher dimensions using a kinematical constraint}} \\[0.5cm]
 
\large{\"{O}zg\"{u}r Akarsu, Tekin Dereli}
\\[0.5cm]

\small{
\textit{Department of Physics, Ko\c{c} University, 34450 Sar{\i}yer, \.{I}stanbul, Turkey}}

\end{center}

\vspace{.6cm}

\hrule \vspace{0.3cm}
\noindent \small{\textbf{Abstract}\\ 
A class of cosmological solutions of higher dimensional Einstein field equations with the energy-momentum tensor of a homogeneous, isotropic fluid as the source are considered with an anisotropic metric that includes the direct sum of a 3-dimensional (physical, flat) external space metric and an $n$-dimensional (compact, flat) internal space metric. A simple kinematical constraint is postulated that correlates the expansion rates of the external and internal spaces in terms of a real parameter $\lambda$. A specific solution for which both the external and internal spaces expand at different rates is given analytically for $n=3$. Assuming that the internal dimensions were at Planck length scales when the external space starts with a Big Bang ($t=0$), they expand only $1.49$ times and stay at Planck length scales even in the present age of the universe ($13.7$ Gyr). The effective four dimensional universe would exhibit a behavior consistent with our current understanding of the observed universe. It would start in a stiff fluid dominated phase and evolve through radiation dominated and pressureless matter dominated phases, eventually going into a de Sitter phase at late times.}
\\
\noindent
\hrule
\def\thefootnote{\arabic{footnote}}
\setcounter{footnote}{0}

\let\thefootnote\relax\footnote{\textbf{E-Mail:} oakarsu@ku.edu.tr, tdereli@ku.edu.tr}

\def\thefootnote{\arabic{footnote}}
\setcounter{footnote}{0}

\section{Introduction}
\label{Intro}
There are neither a priori nor observational reasons for assuming
that the universe during its dynamical evolution has always been
four dimensional. The unification of fundamental interactions of
nature achieved in higher dimensions provides a strong motivation to
give a serious consideration to this possibility. The first attempt
to unify gravitation and electromagnetism by Kaluza and Klein was
based on the idea that the universe we live in is in fact five
dimensional, but as the fifth dimension remains small, it appears
effectively four dimensional \cite{OverduinWesson97}. We know today
that anomaly-free superstring models of all fundamental interactions
require a spacetime of ten dimensions for consistency and the
M-theory in which they are supposedly be embedded lives in an eleven
dimensional spacetime (see \cite{Lidsey00} and references therein).
It is generally assumed that all but four of the spacetime
dimensions are compactified on an unobservable internal manifold,
leaving an observable (1+3)-dimensional spacetime. In the early
1980's the dynamical reduction of internal dimensions to
unobservable scales with the physical, external dimensions
expanding while the internal dimensions contracting, has been
considered for the first time in cosmology
\cite{ChodosDetweiler80,Freund82,DereliTucker83}. Much later
cosmological models where the internal dimensions are static and
remain at unobservable scales while the external space keeps
expanding were also investigated (see for example,
\cite{Bringmann03}). We would like to point out here that there is
yet another possibility. Both of the external and internal
dimensions may start at comparable small scales, yet at later stages
of the evolution of the universe the scale of the internal
dimensions could not expand as fast as that of the external space
does and still remains unobservable. Independent of which
possibility is applied, in a successful higher dimensional
cosmological model, the universe should not only appear effectively
four dimensional today but one should also be able to describe its
dynamical evolution consistently with our present-day observed
universe. The simplest model that fits the present-day cosmological
data is the $\Lambda$-Cold Dark Matter ($\Lambda$CDM) model
\cite{GronHervik}. It is based on Einstein's four-dimensional theory
of general relativity with a spatially flat, isotropic and
homogeneous Robertson-Walker metric. It explains the observed
acceleration of the universe by a simple introduction of a positive
cosmological constant $\Lambda$ that is mathematically equivalent to a
conventional vacuum energy with the equation of state (EoS) parameter
set equal to $-1$. However, this model does not come without any
problems. It suffers from two conceptual problems concerning the
cosmological constant, known as the fine tuning and the coincidence
problems \cite{Carroll92,Copeland06}. The source that drives the observed
acceleration of the universe is still a mystery in the contemporary
cosmology and is usually discussed under the generic name of Dark Energy
(DE). A positive $\Lambda$ is, today, the
simplest candidate for DE besides some scalar field theoretic models
of DE, namely the quintessence, k-essence and others
\cite{Copeland06,Sahni03b}. On the other hand, the dynamics of the observed
universe may be studied in a model independent way known as the
kinematical approach \cite{Rapetti07}. The kinematical approaches to
DE usually favor $w\sim -1$ as well as time-dependent EoS parameters
rather than the constant EoS parameter value $-1$
\cite{Rapetti07,GongWang07,CaiTuo11,Capozziello11}. A
time-dependent EoS parameter is obtained in general, for instance,
when the DE is represented by a scalar field. This is an ad hoc
assumption within four dimensional conventional general relativistic
models. On the other hand, the observed acceleration of the universe can
also be related with the existence of extra space dimensions instead of a
DE field, as will be done here.

In this paper, as the theory of gravitation, we consider the extension of the conventional
four-dimensional Einstein's gravity without $\Lambda$ to higher
dimensions by preserving its mathematical structure. One of the most
important features of unified theories in general is that general
relativity is naturally incorporated in these theories. Such
theories give modifications at very short distances/high energies,
however, they approach Einstein's gravity for sufficiently large
distances/low energies. Hence the use of higher dimensional
Einstein's gravity can also be justified in the context of unified
theories.

\section{The model}

We consider a minimal extension of the conventional
$(1+3)$-dimensional Einstein's field equations to
$(1+3+n)$-dimensions:
\begin{equation}
\label{eqn:EFE} R_{\mu\nu}-\frac{1}{2}Rg_{\mu\nu} =-\kappa
T_{\mu\nu},
\end{equation}
where the indices $\mu$ and $\nu$ run through $0,1,2,...,3+n$ and
$g_{\mu\nu}$, $R_{\mu\nu}$ and $R$ are the metric tensor, the Ricci
tensor and the Ricci scalar, respectively, of a
$(1+3+n)$-dimensional spacetime. $T_{\mu\nu}$ is the
energy-momentum tensor of matter fields in $(1+3+n)$-dimensions and
$\kappa = 8\pi G$ where $G$ is the (positive) gravitational constant
that is to be scaled consistently in $(1+3+n)$-dimensions.

We consider a spatially homogenous but not necessarily isotropic
$(1+3+n)$-dimensional synchronous spacetime metric that involves a
maximally symmetric three dimensional flat external (physical) space
metric and a compact $n$ dimensional flat internal space metric:
\begin{eqnarray}
\label{eqn:metric} ds^2=-dt^2+a^2(t) \left (dx^{2}+dy^{2}+dz^{2}\right)+ s^2(t) \left ( d\theta_{1}^{2} +...+d\theta_{n}^{2}\right).
\end{eqnarray}
$a(t)$ is the scale factor of the external space that represents the
space we observe today while $s(t)$ is the scale factor of the
$n=1,2,3,\dots$ dimensional internal space that cannot be observed
directly and locally today.

We consider the energy-momentum tensor of a $(1+3+n)$-dimensional
homogeneous and isotropic ideal fluid:
\begin{equation}
\label{eqn:EMT} {T^{\mu}}_{\nu}={\textnormal{diag}}[-\rho,
p,p,p,p,...,p],
\end{equation}
where $\rho=\rho(t)$ and $p=p(t)$ are the energy density and
pressure of the fluid.

$(1+3+n)$-dimensional Einstein's field equations (\ref{eqn:EFE}) for
the spacetime described by the metric (\ref{eqn:metric}) in the
presence of a co-moving fluid represented by the energy-momentum
tensor (\ref{eqn:EMT}) read:
\begin{subequations}
\begin{align}
        3\frac{\dot{a}^2}{a^2}+3n\frac{\dot{a}}{a}\frac{\dot{s}}{s}+\frac{1}{2}n(n-1)\frac{\dot{s}^2}{s^2}&=\kappa\rho, \label{eqn:EFE1} \\
        \frac{\dot{a}^2}{a^2}+2\frac{\ddot{a}}{a}+n\frac{\ddot{s}}{s}+2n\frac{\dot{a}}{a}\frac{\dot{s}}{s}+\frac{1}{2}n(n-1)\frac{\dot{s}^2}{s^2}
&= -\kappa p,  \label{eqn:EFE2}\\
3\frac{\dot{a}^2}{a^2}+3\frac{\ddot{a}}{a}+(n-1)\frac{\ddot{s}}{s}+3(n-1)\frac{\dot{a}}{a}\frac{\dot{s}}{s}
+\frac{1}{2}(n-1)(n-2)\frac{\dot{s}^2}{s^2}
&= -\kappa p. \label{eqn:EFE3}
\end{align}
\end{subequations}
This system consists of three differential equations
(\ref{eqn:EFE1})-(\ref{eqn:EFE3}) that should be satisfied by four
unknown functions $a$, $s$, $\rho$, $p$ and therefore is not fully
determined. It is customary at this point either to introduce an
equation of state that characterizes the internal properties of the
fluid or alternatively to make a kinematical ansatz to fully
determine the system. However, even in four dimensional
accelerating cosmological models the choice of the DE fluid is ad
hoc. In our case, we almost have no clue concerning the nature of a
possible higher dimensional fluid. Hence, we find it natural rather
to postulate an ansatz that correlates the kinematics between the external and internal spaces to fully determine the system. In the field equations (\ref{eqn:EFE1})-(\ref{eqn:EFE3}), the external and internal dimensions couple directly through the term
\begin{equation}
\frac{\dot{a}}{a}\frac{\dot{s}}{s}=f(t),
\end{equation}
which most generally will be a function of the cosmic time $t$. We note that $f(t)$ is determined by the kinematics of both the external and internal spaces and hence in return one can correlate the kinematics of the internal and external spaces by specifying a function for $f(t)$ and can characterize the properties of the higher dimensional cosmology. For an expanding external space $\frac{\dot{a}}{a}>0$ and therefore the positive values of $f(t)$ correspond to an expanding internal space, while the negative values of $f(t)$ correspond to a contracting internal space. On the other hand, $f(t)=0$ describes the Kaluza-Klein reduction, i.e., one will obtain a cosmological solution in which the internal space is static. In this work we are particularly interested in the possibility of viable higher dimensional cosmological models in which both the external and internal spaces are expanding so that $f(t)>0$. In line with the above discussion, to determine the field equations fully, we propose the simplest generalization of the case $f(t)=0$ for which
\begin{equation}
\label{eqn:constraint}
\frac{\dot{a}}{a}\frac{\dot{s}}{s}=\frac{\lambda}{9},
\end{equation}
where $\lambda$ is a real constant. Since the fluid is isotropic we
eliminate the pressure between (\ref{eqn:EFE2})-(\ref{eqn:EFE3}),
and use the resulting equation together with (\ref{eqn:constraint})
to solve for the scale functions $a$ and $s$. Then we substitute
these in (\ref{eqn:EFE1}) and (\ref{eqn:EFE2}) to get $\rho$ and
$p$, respectively. We were not able to get analytical expressions
for arbitrary values of $n$. Therefore we give explicit solutions
below only for $n=3$ (Numerical solutions might be studied for other values of $n$):

\begin{equation}
\label{eqn:gensola}
        a=a_{0}t^{\frac{1}{3}}\quad \textnormal{and}\quad s=s_{0} \quad \textnormal{for} \quad \lambda=0,
\end{equation}
where $a_{0}$ and $s_{0}$ are constants of integration, and
\begin{equation}
\label{eqn:gensol}
a=\left(c_{1}e^{\sqrt{\lambda}\,t}-c_{2}e^{-\sqrt{\lambda}\,t}\right)^{\frac{1}{3}}\quad \textnormal{and}\quad s=c_{3}\left(c_{1}e^{\sqrt{\lambda}\,t}+c_{2}e^{-\sqrt{\lambda}\,t}\right)^{\frac{1}{3}}\quad \textnormal{for}\quad \lambda\neq 0,
\end{equation}
where $c_{1}$, $c_{2}$ and $c_{3}$ are constants of integration.
One may check that, depending on the choice of the integration
constants and $\lambda$, the scale factors exhibit five different
types of behavior\footnote{We would like to note that kinematics similar to that we obtained for the external space for $\lambda\neq 0$ is also noted by Capozziello et al. \cite{Capozziello97}, although with a totally different reasoning in the context of conventional, four dimensional relativistic cosmology.}:
\begin{enumerate}[(i)]
\item
$\lambda=0$: The external space expands as in the four dimensional
universe that is filled with a stiff fluid\footnote{Stiff fluid is the most promising EoS of matter at ultra-high densities for representing the very early universe (see \cite{Zeldovich62,Barrow78}) and is described with an EoS parameter $p/\rho=1$, where $\rho$ and $p$ are the energy density and pressure, respectively.}, while the internal space is
static.
\item
$\lambda>0$ and $c_{1}\neq0= c_{2}$: Both of the external and
internal spaces expand exponentially at the same rate.
\item
$\lambda>0$ and $c_{1}=0\neq c_{2}$: Both of the external and
internal spaces contract exponentially at the same rate.
\item
$\lambda>0$ and $c_{1}\neq0\neq c_{2}$: The scale functions can be
written in terms of hyperbolic functions.
\item
$\lambda<0$ and $c_{1}\neq0\neq c_{2}$: The scale functions can be
written in terms of sinusoidal functions.
\end{enumerate}
 In what follows, we
concentrate in particular on the case (iv) with the additional
condition $c_{1}c_{2}>0$. We will show that the external space
exhibits a $\Lambda$CDM-type behavior, while the internal space
expands at a much slower rate than the external space.

\section{An effective four dimensional $\Lambda$CDM-type cosmology}

\subsection{Solution of the higher dimensional equations}

It is easy to check that for $c_{1}c_{2}>0$ and $\lambda>0$, the
scale factor of the external space is null $a=0$ at
$t=\frac{1}{2\sqrt{\lambda}}\ln{\left(\frac{c_{2}}{c_{1}}\right)}$.
Hence, for convenience, we may set the singularity of the external
space at $t=0$ with the choice $c_{1}=c_{2}$ without loss of
generality\footnote{If $c_{1}c_{2}>0$, in the case $c_{1}\neq c_{2}$
the evolution of the Hubble and deceleration parameters turn out to
be exactly the same with the ones in the case $c_{1}=c_{2}$, but
shifted along the time axis.}. Choosing $c_{1}=c_{2}$ in
(\ref{eqn:gensol}) and re-naming the integration constants, we
obtain the cosmological parameters of the external dimensions; the
scale factor, Hubble parameter and deceleration parameter,
respectively, as follows:

\begin{subequations}
\begin{align}
        a&=a_{1}\sinh^{\frac{1}{3}}(\sqrt{\lambda}\,t),\\
        H_{a}=\frac{\dot{a}}{a}&=\frac{\sqrt{\lambda}}{3}\coth(\sqrt{\lambda}\,t),\\
        q_{a}=-\frac{\ddot{a} a}{\dot{a}^2}&=-1+3\, {\rm sech}^{2}(\sqrt{\lambda}\,t),
\end{align}
\end{subequations}
and of the internal dimensions, respectively, as follows:
\begin{subequations}
\begin{align}
        s&=s_{1}\cosh^{\frac{1}{3}}(\sqrt{\lambda}\,t),\\
        H_{s}=\frac{\dot{s}}{s}&=\frac{\sqrt{\lambda}}{3}\tanh(\sqrt{\lambda}\,t),\\
       q_{s}=-\frac{\ddot{s} s}{\dot{s}^2}&=-1-3\, {\rm cosech}^{2}(\sqrt{\lambda}\,t),
\end{align}
\end{subequations}
where $a_{1}$ and $s_{1}$ are the new integration constants. The
energy density, pressure and EoS parameter of the higher dimensional
fluid are given, respectively, as follows:

\begin{subequations}
\label{eqn:7Dfluid}
\begin{align}
        \rho&= \frac{4 \lambda}{3 \kappa}\,{{\rm
cosech}^{2}(2\sqrt{\lambda}\,t)}+ \frac{5 \lambda}{3 \kappa},  \\
        p&=\frac{4 \lambda}{3 \kappa}\,{{\rm cosech}^{2}(2\sqrt{\lambda}\,t)}-\frac{5 \lambda}{3\kappa},\\
       w&=\frac{p}{\rho}=\frac{{4 - 5 \sinh^{2}(2\sqrt{\lambda}\,t)}}{{4 +
5\sinh^{2}(2\sqrt{\lambda}\,t)}}.
\end{align}
\end{subequations}

\begin{figure}[ht]
\begin{minipage}[b]{0.49\linewidth}
\centering
\includegraphics[width=1\textwidth]{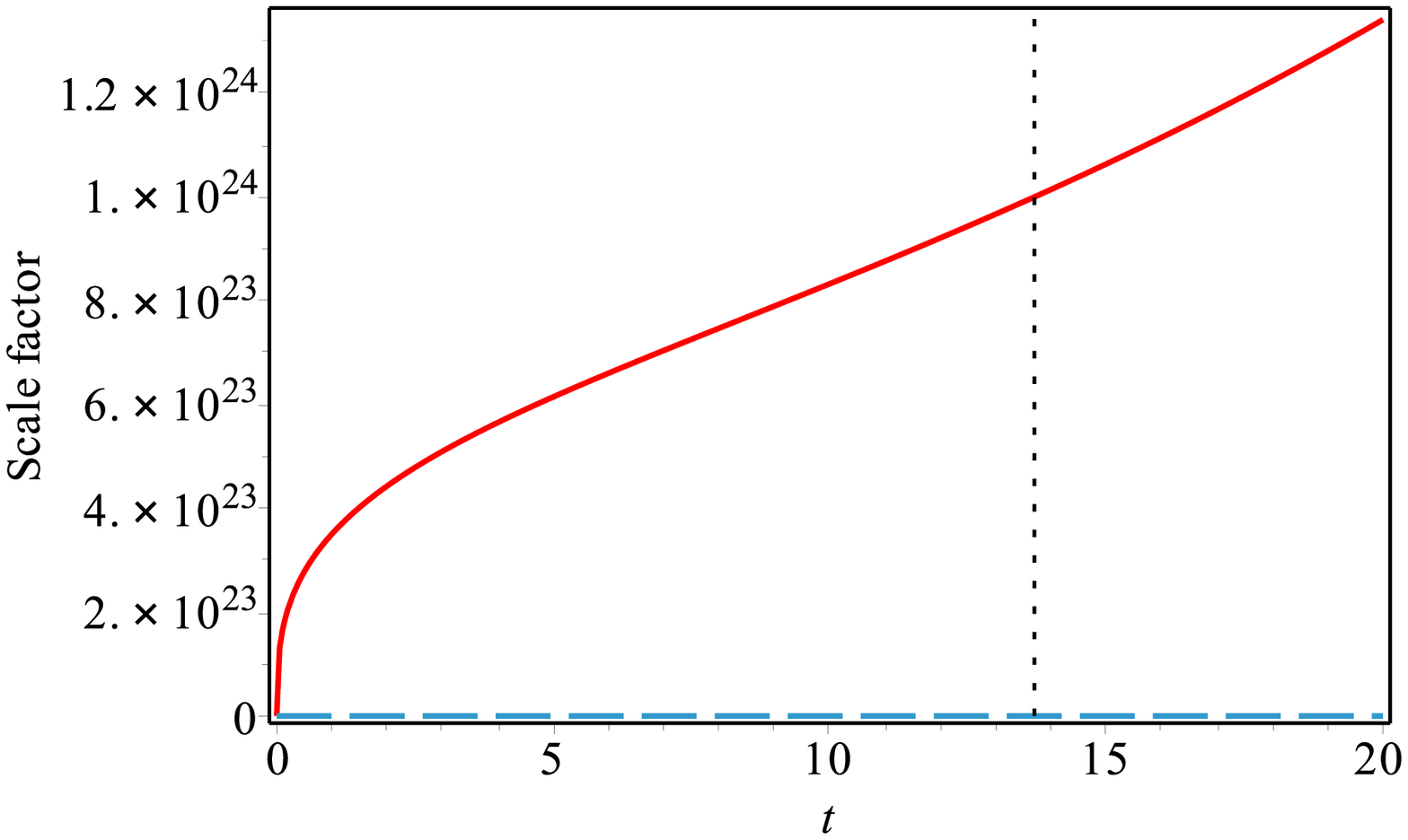}
\caption{The scale factors (meter) of the external (solid) and internal (dashed)
dimensions vs. cosmic time $t$ (Gyr). At $t=0$, the external dimensions are null and internal dimensions are at Planck length scale.}
\label{fig:sfs}
\end{minipage}
\hspace{0.01\linewidth}
\begin{minipage}[b]{0.49\linewidth}
\centering
\includegraphics[width=1\textwidth]{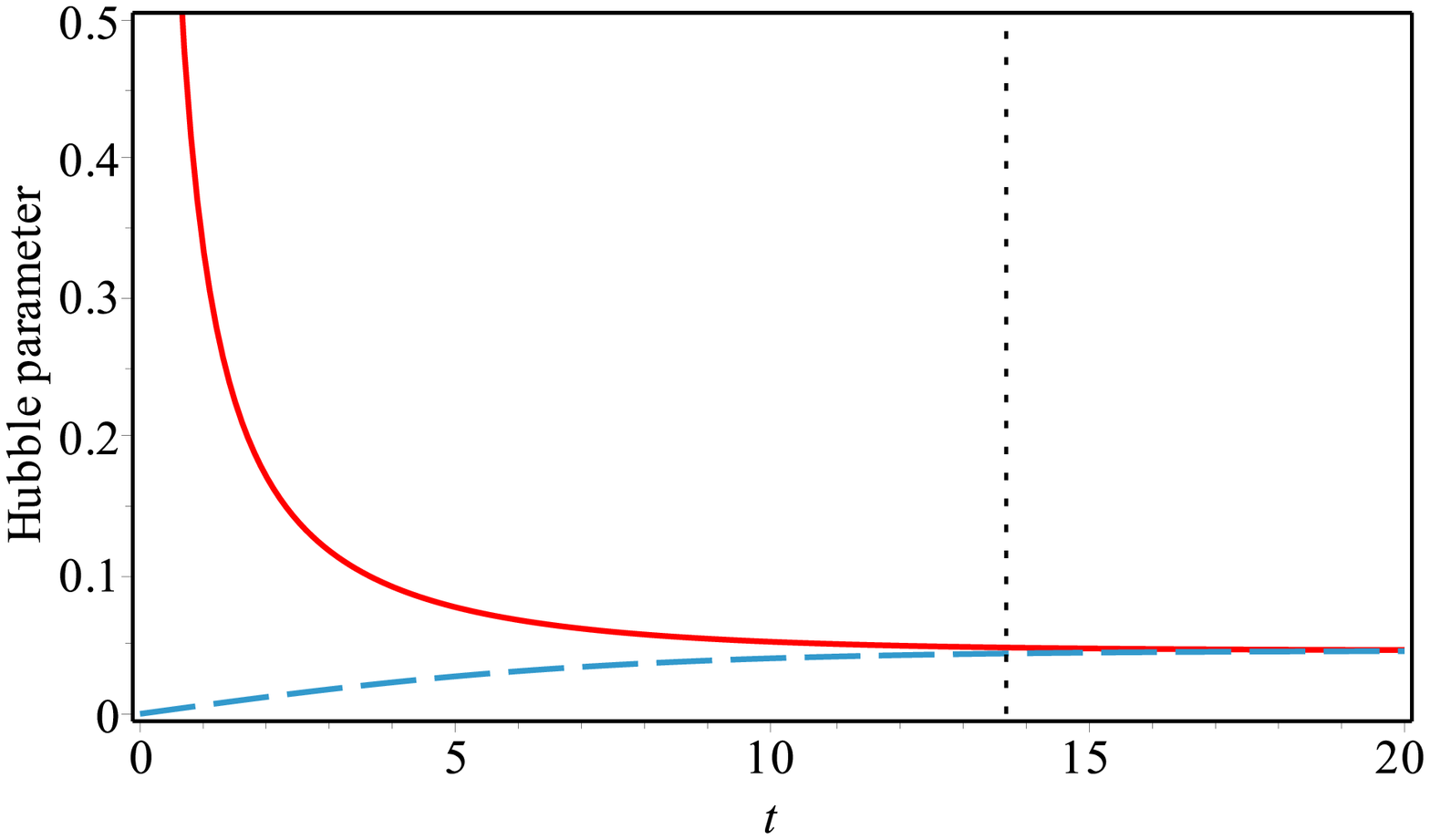}
\caption{The Hubble parameters of the external (solid) and internal (dashed) dimensions vs. cosmic time $t$ (Gyr). The expansion rate of the external space is always higher than of the internal space.}
\label{fig:hps}
\end{minipage}
\end{figure}

It may be seen from the expressions above that both of the external
and internal spaces expand for $t>0$. However, at the instant $t=0$,
while the external space starts expanding from zero $(a=0$) with
an infinitely large expansion rate ($H_{a}=\infty$ and $q_{a}=2$);
the internal space will be static ($H_{s}=0$ and $q_{s}=\infty$)
remaining at a non-zero size $s=s_{1}$. Indeed, when the scale
factors are Taylor expanded
\begin{subequations}
\begin{align}
        a&=a_{1}{\lambda}^{\frac{1}{6}}\,t^{\frac{1}{3}}+a_{1}\frac{{\lambda}^{\frac{7}{6}}}{18}\,t^{\frac{7}{3}}
+O(t^{\frac{13}{3}}), \\
        s&=s_{1}+s_{1}\frac{{\lambda}}{6}\,t^{2}+O(t^{4}),
\end{align}
\end{subequations}
we see that $a\sim t^{\frac{1}{3}}$ while $s\sim s_{1}$ as $t\sim0$;
that is, in the very early times of the expansion, the external
space volume $a^3$ grows almost linearly with time, while the
internal space volume $s^3$ is almost constant (see Fig.
\ref{fig:sfs}). Furthermore one may check that the expansion rate of
the internal dimensions is always smaller than that of the external
dimensions during the entire history of the universe i.e.,
$H_{a}>H_{s}$, and they approach each other in the infinite future,
i.e., $H_{a}\rightarrow \frac{\sqrt{\lambda}}{3}$ and
$H_{s}\rightarrow \frac{\sqrt{\lambda}}{3}$ as $t\rightarrow \infty$
(see Fig. \ref{fig:hps}). 
Hence, if the internal dimensions start to
expand at an unobservable length scale (for instance, at $s_{1}\sim
l_{{\rm{Planck}}}\sim10^{-35}$ m), they might not be able to expand
to observable length scales (say for instance, to $\sim 10^{-20}$ m
which is the scale that corresponds to the energy scale of $\rm TeV$
that is probed by the Large Hadron Collider (LHC)) even today. In
the mean time, the external dimensions will expand from its initial
singularity to its present-day observed length scale ($10^{24}$ m).
Both the external and internal dimensions would have grown from
their minimal values $a=0$ and $s=s_{1}$ at $t=0$ to an equal size
at time
\begin{equation}
t_{{\rm{eq}}}=\frac{1}{2\sqrt{\lambda}}\ln{\left(\frac{a_{1}^3+s_{1}^3}{a_{1}^3-s_{1}^3}\right)}.
\end{equation}
Therefore, if $s_{1}\sim l_{planck}\ll a_{1}$ one can safely take
$t_{{\rm{eq}}}\sim 0$. We may determine how many times the sizes of
the external and internal dimensions expanded since the time
$t_{{\rm{eq}}}$ when they were equal:
\begin{subequations}
\begin{align}
        \frac{a(t)}{a(t_{{\rm{eq}}})}&=\frac{\sinh^{\frac{1}{3}}(\sqrt{\lambda}\;t)}{\sinh^{\frac{1}{3}}\left(\frac{1}{2}\ln{\left(\frac{a_{1}^3+s_{1}^3}{a_{1}^3-s_{1}^3}\right)}\right)}, \\
        \frac{s(t)}{s(t_{{\rm{eq}}})}&=\frac{\cosh^{\frac{1}{3}}(\sqrt{\lambda}\;t)}{\cosh^{\frac{1}{3}}\left(\frac{1}{2}\ln{\left(\frac{a_{1}^3+s_{1}^3}{a_{1}^3-s_{1}^3}\right)}\right)}.
\end{align}
\end{subequations}
The choice $s_{1}\ll a_{1}$ implies
$\frac{a(t)}{a(t_{{\rm{eq}}})}\gg\frac{s(t)}{s(t_{{\rm{eq}}})}$ for
all $t\gg t_{{\rm{eq}}}$. It is also interesting to note that how many
times the size of the internal dimensions have grown compared to
their initial size may be determined just by the
 present-day value of the deceleration parameter of the external
dimensions. To show this, we simply isolate $\lambda$ in $q_{a}(t)$
and substitute it in $s(t)$ above and obtain:
\begin{equation}
\label{eqn:intfold}
\frac{s}{s_{1}}=\left(\frac{3}{q_{a}+1}\right)^{\frac{1}{6}}
\end{equation}
which gives us the ratio $\frac{s}{s_{1}}$ for any given value of
$q_{a}$. Hence one can easily calculate how many times the size of
the internal dimensions have grown since the beginning of time to
the present-day simply by measuring the present-day value of the
deceleration parameter of the observed universe. Using $q_{a}=-0.73$
\cite{Cunha09} for the present-day value of the dimensionless
deceleration parameter of the external space and setting
$t_{0}=13.7$ (Gyr) for the present age of the universe we obtain
$\lambda=0.0187$. We take the present size of the visible universe
as $10^{24}$ m and going backwards obtain the value
$a_{1}=6.8\times 10^{23}$ m. If we now assume that the internal
dimensions were at Planck length scales at time $t=0$, i.e., $s_{1}=
l_{\rm Planck}\sim 10^{-35}$ m, then the external and internal
dimensions would have reached the same size when $t_{{\rm{eq}}}=
2.32 \times 10^{-176}$ (Gyr). The external dimensions will expand
$\frac{a(13.7)}{a(t_{{\rm{eq}}})}\simeq 10^{59}$ times during the
time interval $13.7-t_{{\rm{eq}}}$ (Gyr) while the internal dimensions
expand only $\frac{s(13.7)}{s(t_{{\rm{eq}}})}\simeq 1.49$ times! The
same conclusion for the internal dimensions may be reached simply
by using $q_{a}=-0.73$ in (\ref{eqn:intfold}) so that
$\frac{s}{s_{1}}\simeq 1.49$.

On the other hand, the internal dimensions expand from the $l_{\rm
Planck}$ length scales at the beginning to the LHC length scales
($10^{-20}$ m) at $t=763$ (Gyr), the proton size ($10^{-15}$ m) at
$t=1015$ (Gyr) and the meter length scales at $t=1773$ (Gyr). In
short, according to our model, all the dimensions that were at
Planck scales $l_{\rm Planck}$ at time $t_{{\rm{eq}}}= 2.32 \times
10^{-176}$ (Gyr) evolve in such a way that the external dimensions
are today at length scales $10^{24}$ m while the internal
dimensions are still at Planck length scales $l_{\rm Planck}$ (see
Fig. \ref{fig:sfs}).

Our model also predicts that the present value of the deceleration
parameter of the observed universe must be strictly higher than -1,
i.e. $q_{a}> -1$, otherwise we would have observed the extra
dimensions, since $\frac{s}{s_{1}}\rightarrow \infty$ as
$q_{a}\rightarrow -1$.

Finally, the energy density and pressure of our higher dimensional
ideal fluid will be infinitely large at the beginning. They decrease
monotonically and approach $\rho\sim \frac{5 \lambda}{3 \kappa}$ and
$p\sim -\frac{5 \lambda}{3 \kappa}$, respectively, for sufficiently
large values of $t$. The EoS parameter of the fluid, on the other hand,
starts with $w=1$ at $t=0$ and approaches $w\sim-1$ for sufficiently
large $t$ values. We won't be dwelling on the properties of this
higher dimensional fluid further, however, its manifestations in the
effective four dimensional universe will be discussed below.

The above calculations show that, although the internal dimensions
are also expanding just as the (observable) external dimensions do,
they remain far too small to allow for local and direct detection
today and in the near future. However, their presence obviously has tremendous effect on
our cosmological history. We have here a durable model of the
effective four dimensional universe. But this is not yet enough. We
should further investigate whether this predicted effective four
dimensional universe is consistent with the present-day cosmological
observations. We shall deal with this question in the following subsection.

\subsection{The effective four dimensional universe}

\begin{figure}[h]
\centering
\includegraphics[width=0.50\textwidth]{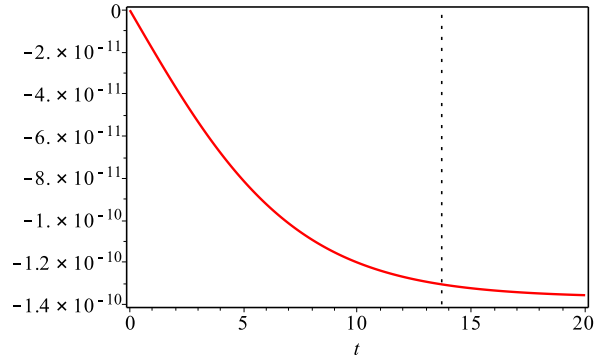}
\caption{The time (yr) variation of the 4-dimensional effective gravitational coupling ($\dot{\tilde{\kappa}}/\tilde{\kappa}\;{\rm yr}^{-1}$) vs. cosmic time $t$ (Gyr).}
\label{fig:G4var}
\end{figure}

\begin{figure}[h]

\begin{minipage}[b]{0.49\linewidth}
\centering
\includegraphics[width=1\textwidth]{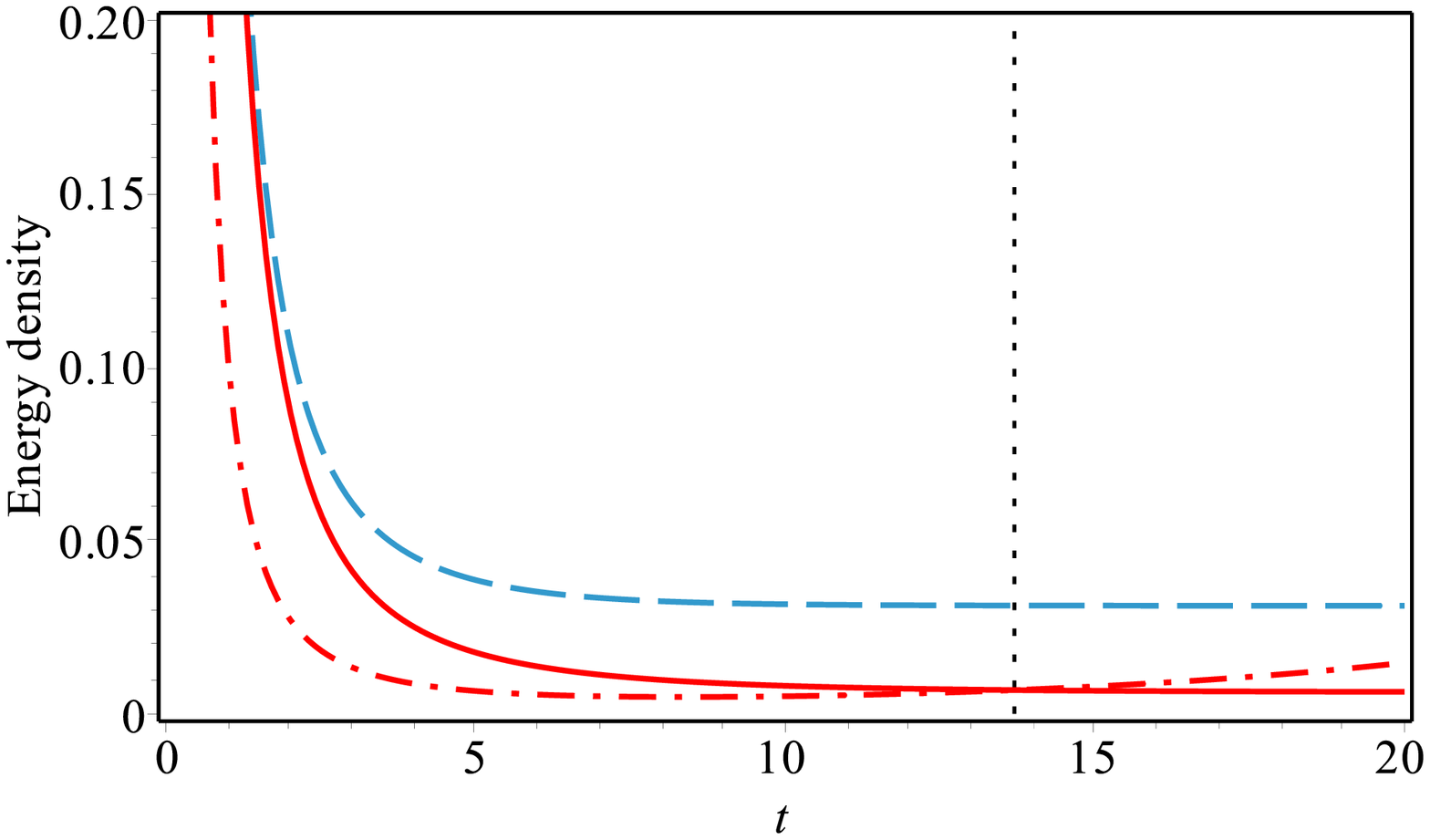}
\caption{The energy densities of the 4-dimensional effective fluids (solid for $\tilde{T}_{ij}$ and dashed-dotted for $\tilde{T}'_{ij}$) and the higher dimensional fluid (dashed) vs. cosmic time $t$ (Gyr).}
\label{fig:rhoNEW}
\end{minipage}
\hspace{0.01\linewidth}
\begin{minipage}[b]{0.49\linewidth}
\centering
\includegraphics[width=1\textwidth]{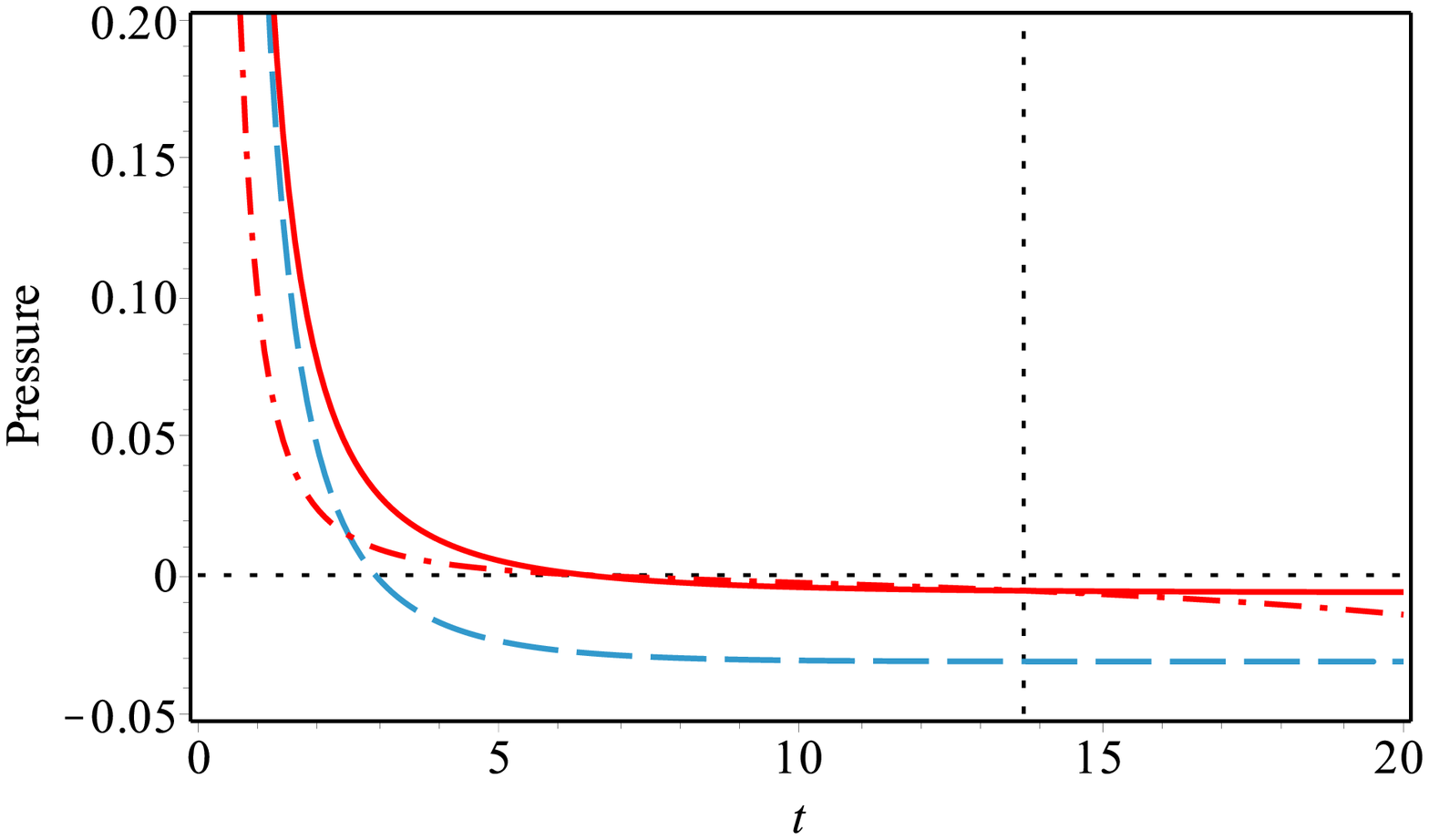}
\caption{The pressures of the 4-dimensional effective fluids (solid for $\tilde{T}_{ij}$ and dashed-dotted for $\tilde{T}'_{ij}$) and the higher dimensional fluid (dashed) vs. cosmic time $t$ (Gyr).}
\label{fig:pNEW}
\end{minipage}

\end{figure}

\begin{figure}
\centering
\includegraphics[width=0.5\textwidth]{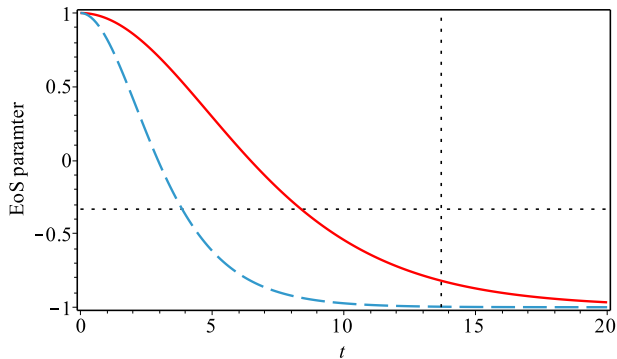}
\caption{The equation of state parameters (EoS) of the four (solid) and higher (dashed) dimensional effective fluids vs. cosmic time $t$ (Gyr). EoS parameter of the four dimensional effective fluid is $-\frac{1}{3}$ at $t=8.38$ (Gyr).}
\label{fig:w4d}
\end{figure}

In cosmology, we do not usually deal with direct measurements of the
energy density and pressure of the material/physical content of the
universe. We collect data on the kinematics of the observed universe
instead, e.g., from the supernova Ia observations
\cite{Rapetti07,GongWang07,CaiTuo11,Capozziello11,Capozziello97,Cunha09} and on the geometry of the
space from cosmic microwave background by WMAP observations
\cite{Komatsu11}. Furthermore, we assume that the space we live in
is (effectively) three dimensional. Then, what we do in general is
to interpret the collected information using a reliable theory, for
instance the general relativity of Einstein, to infer the properties
of the material content of the universe. This is, naturally, the
approach of an observer who is unaware of internal dimensions. On
the other hand, we had been arguing all along that we may in fact be
living in a higher dimensional space which appears effectively three
dimensional since the internal dimensions are today so small that
they evade direct and local detection. However, the internal
dimensions may still be controlling the dynamics of the external
dimensions that we observe. Hence, while we are interpreting the
cosmological data within the framework of four dimensional general
relativity, the components related to the internal dimensions and
the higher dimensional fluid we introduced could manifest themselves
as an effective source in the 4-dimensional Einstein's field
equations. An observer who lives in four dimensions would naturally
use the 4-dimensional Einstein's field equations:
\begin{equation}
\label{eqn:Einr} \tilde{R}_{ij}-\frac{1}{2}\tilde{R}\tilde{g}_{ij}
=-\tilde{\kappa}_{0} \tilde{T}_{ij},
\end{equation}
where $i$ and $j$ run through $0,1,2,3$ and $\tilde{\kappa}_{0}=8\pi
\tilde{G}_{0}$ with $\tilde{G}_{0}$ being the value of the four dimensional gravitational coupling that is observed with local experiments today. $\tilde{R}_{ij}$, $\tilde{R}$ and $\tilde{g}_{ij}$
are the Ricci tensor, Ricci scalar and the metric tensor of the
($1+3$)-dimensional spacetime, respectively. $\tilde{T}_{ij}$
refers to the components of the four dimensional effective
energy-momentum tensor. In the 4-dimensional spatially flat RW
spacetime, effective Einstein field equations read:
\begin{subequations}
\begin{align}
        3\frac{\dot{a}^2}{a^2}&=\tilde{\kappa}_{0}\tilde{\rho}, \label{eqn:EFE4d1} \\
       \frac{\dot{a}^2}{a^2}+2\frac{\ddot{a}}{a}&= -\tilde{\kappa}_{0}\tilde{p}. \label{eqn:EFE4d2}
\end{align}
\end{subequations}
A comparison of these equations with the higher dimensional field
equations given before, leads to the following identifications:
\begin{subequations}
\begin{align}
       \tilde{\rho}&=\frac{\kappa}{\tilde{\kappa}_{0}}\rho-\frac{\lambda}{\tilde{\kappa}_{0}}-
\frac{3 \dot{s}^2}{\tilde{\kappa}_{0} s^2},\\
      \tilde{p}&=\frac{\kappa}{\tilde{\kappa}_{0}}p+ \frac{3
\ddot{s}}{\tilde{\kappa}_{0}s}+\frac{2 \lambda}{3 \tilde{\kappa}_{0}}+
\frac{3 \dot{s}^2}{\tilde{\kappa}_{0} s^2}.
\end{align}
\end{subequations}
One may now observe how the components of the higher dimensional
distributions manifest themselves in an effective energy-momentum source in the four
dimensional universe. Also note that although an observer cannot
observe the internal dimensions directly and locally, the internal
dimensions contribute in an essential way to the dynamics of the
external dimensions. Substituting $a$ into the four dimensional
field equations (\ref{eqn:EFE4d1}) and (\ref{eqn:EFE4d2}), the observer would obtain the
energy density, pressure and the EoS parameter of the observed universe
as follows:
\begin{subequations}
\begin{align}
        \tilde{\rho}&=\frac{\lambda}{3\tilde{\kappa}_{0}}{\rm
cosech}^{2}(\sqrt{\lambda}\,t)+\frac{\lambda}{3\tilde{\kappa}_{0}}, \label{eqn:4Dfluid1}\\
       \tilde{p}&=\frac{\lambda}{3\tilde{\kappa}_{0}}{\rm
cosech}^{2}(\sqrt{\lambda}\,t)-\frac{\lambda}{3\tilde{\kappa}_{0}}, \label{eqn:4Dfluid2}\\
\tilde{w}&=\frac{\tilde{p}}{\tilde{\rho}}=\frac{1
-\sinh^{2}(\sqrt{\lambda}\,t)}{1+\sinh^{2}(\sqrt{\lambda}\,t)}. \label{eqn:4Dfluid3}
\end{align}
\end{subequations}
These are the properties of a 4-dimensional effective fluid that are inferred by an observer who is interpreting the kinematics of the observed universe through the 4-dimensional conventional general relativity, in which the gravitational coupling is a constant $\tilde{\kappa}_{0}$. However, in a higher dimensional universe, even when the internal space remains at an unobservable size the gravitational field will be propagating in the full higher dimensional space and hence the strength of the 4-dimensional effective gravitational coupling $\tilde{\kappa}=8\pi \tilde{G}$ will be related to the higher dimensional gravitational coupling constant through the proper volume of the internal space $V^{\rm int}\propto s^3$ as follows \cite{Dvali99,Uzan11}:
\begin{equation}
\label{eqn:G4}
\tilde{\kappa}=\frac{\kappa}{V^{\rm int}}.
\end{equation}
Accordingly, the dynamics of the internal space may manifest itself by giving rise to a time variation of the 4-dimensional effective gravitational coupling. Let us now check whether the time variation of $\tilde{\kappa}$ is consistent with the observational constraints and whether it is possible for the observer to detect how $\tilde{\kappa}$ varies in time. Using (\ref{eqn:G4}) we obtain
\begin{equation}
\label{eqn:G4var1}
\tilde{\kappa}=\tilde{\kappa}_{0}\frac{V^{\rm int}_{0}}{V_{\rm int}}=\tilde{\kappa}_{0}\frac{s^{3}_{0}}{s^{3}},
\end{equation}
which gives
\begin{equation}
\frac{\dot{\tilde{\kappa}}}{\tilde{\kappa}}=-3H_{s}=-\sqrt{\lambda}\tanh(\sqrt{\lambda}t)
\end{equation}
for the time variation of $\tilde{\kappa}$. We immediately notice that the time variation of the 4-dimensional gravitational coupling is null at $t=0$, decreases with the cosmic time $t$ and approaches $-\sqrt{\lambda}$ as $t\rightarrow \infty$. Using $\lambda=0.0187$ (Gyr$^{-2}$) we find that the time variation of $\tilde{\kappa}$ is null at $t=0$, $\sim -10^{-25}\,{\rm yr}^{-1}$ at $t\sim 10^{2}\,{\rm s}$ (the time scale of the primordial nucleosynthesis in the standard model for the history of the universe), $\sim -10^{-15}\,{\rm yr}^{-1}$ at $t\sim 10^{5}\,{\rm yr}$ (the time scale of photon decoupling in the standard model for the history of the universe), $-1.3\times 10^{-10}\,{\rm yr}^{-1}$ at the present age of the universe and goes to $-1.4\times 10^{-10}\,{\rm yr}^{-1}$ as $t\rightarrow \infty$. The time variation of $\tilde{\kappa}$ is plotted in Fig. \ref{fig:G4var}.
We also calculated the average value of the time variation of $\tilde{\kappa}$ from $t=0$ to the present age of the universe $13.7$ (Gyr): 
\begin{equation}
\frac{1}{13.7\times 10^{9}\,{\rm yr}}\int_{t=0}^{t=13.7\times 10^{9} {\rm yr}}\frac{\dot{\tilde{\kappa}}}{\tilde{\kappa}} {\rm d}t=-8.8\times 10^{-11}\,{\rm yr}^{-1}.
\end{equation}
The majority of constraints on the time variation of $\tilde{\kappa}$ coming from the Solar system, pulsar timing or stellar observations that is found in the literature favor a value $\sim \pm 10^{-11}\,{\rm yr}^{-1}$ in the vicinity of the present age of the universe \cite{Uzan11}. Considering the random and the systematic errors involved in the determination of such constraints, the time variation of $\tilde{\kappa}$ in our model is consistent with the above value. The most severe constraints in the literature are set from primordial nucleosynthesis and imply that when the primordial nucleosynthesis took place in the early universe it was $\sim 10^{-12}\,{\rm yr}^{-1}$, which is also in line with our very small value $\sim -10^{-25}\,{\rm yr}^{-1}$ for the time scale of that epoch. On the other hand, considering the very small values of these constraints, it will be natural for the observer to conceive the 4-dimensional gravitational coupling as a constant. In the absence of any information of the presence of internal dimensions or the time variation of the 4-dimensional gravitational coupling, an observer would conclude that the expansion of the observed universe is governed by an unknown "dark energy" source whose properties are given by (\ref{eqn:4Dfluid1})-(\ref{eqn:4Dfluid3}). Let us now suppose that the observer is able to resolve the time variation of $\tilde{\kappa}$ correctly from observations. In this case, using (\ref{eqn:Einr}) and (\ref{eqn:G4var1}), one can define a new 4-dimensional effective energy-momentum tensor $\tilde{T}'_{ij}$ for the 4-dimensional effective fluid that is related to $\tilde{T}_{ij}$ as follows:
\begin{equation}
\tilde{T}'_{ij}=\frac{s^{3}}{s^{3}_{0}}\tilde{T}_{ij}.
\end{equation}
Hence
\begin{eqnarray}
\tilde{\rho}'=\frac{s^{3}}{s_{0}^{3}}\tilde{\rho}, \quad \tilde{p}'=\frac{s^{3}}{s_{0}^{3}}\tilde{p} \quad\textnormal{and}\quad \tilde{w}'=\frac{\tilde{p}'}{\tilde{\rho}'}=\tilde{w}
\end{eqnarray}
for the energy density, pressure and the EoS of the 4-dimensional effective fluid respectively. We plot the evolution of the 4-dimensional effective and higher dimensional energy densities in Fig. \ref{fig:rhoNEW}, pressures in Fig. \ref{fig:pNEW} and EoS parameters in Fig. \ref{fig:w4d}. 
It is worth noting that in this case the 4-dimensional theory of gravitation is not conventional general relativity anymore; it is a theory that yields the same mathematical form with general relativity but involves a time dependent gravitational coupling. We note that the EoS parameter of the 4-dimensional effective fluid remains the same. The energy densities and pressures of these two energy-momentum tensors coincide today and remain almost the same in the vicinity of the present age of the universe. On the other hand, they differ slightly at earlier times of the universe and will differ considerably in the far future. Note however that the universe also can no longer be taken as effectively 4-dimensional in the far future since the internal dimensions grow considerably in size. Hence, because both the internal dimensions and the time variation of the 4-dimensional gravitational coupling is out of the reach of the observer, it is a very good approximation for the observer to interpret the expansion of the observed universe through the conventional 4-dimensional general relativity which describes the local gravitational events (in the Solar system) successfully.

\begin{figure}[ht]

\begin{minipage}[b]{0.49\linewidth}
\centering
\includegraphics[width=1\textwidth]{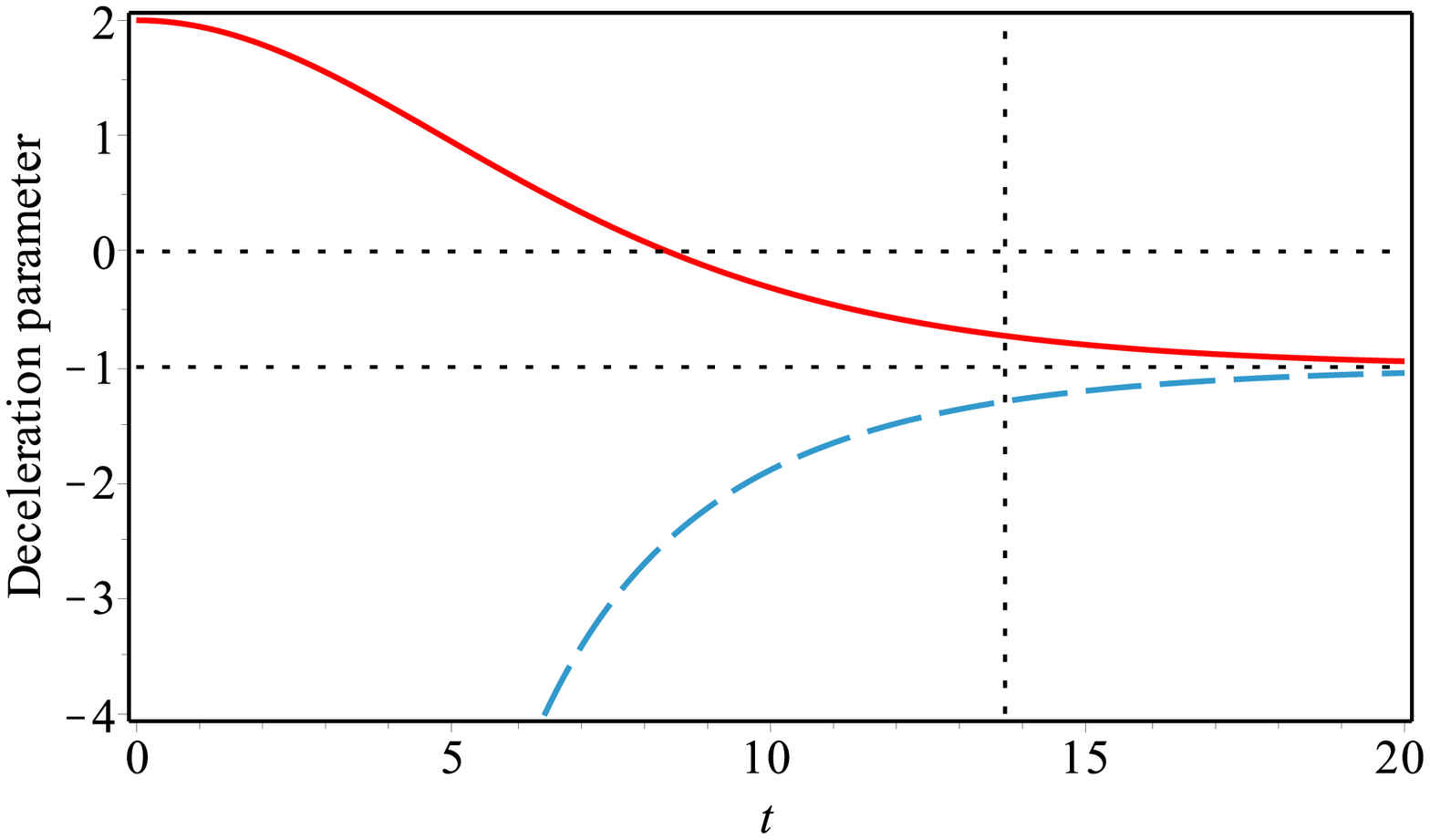}
\caption{The deceleration parameters of the external (solid) and internal (dashed) dimensions vs. cosmic time $t$ (Gyr). The external dimensions start accelerating at $t_{\rm t}=8.38$ (Gyr), i.e., $5.32$ (Gyr) ago from today.}
\label{fig:dps}
\end{minipage}
\hspace{0.01\linewidth}
\begin{minipage}[b]{0.49\linewidth}
\centering
\includegraphics[width=1\textwidth]{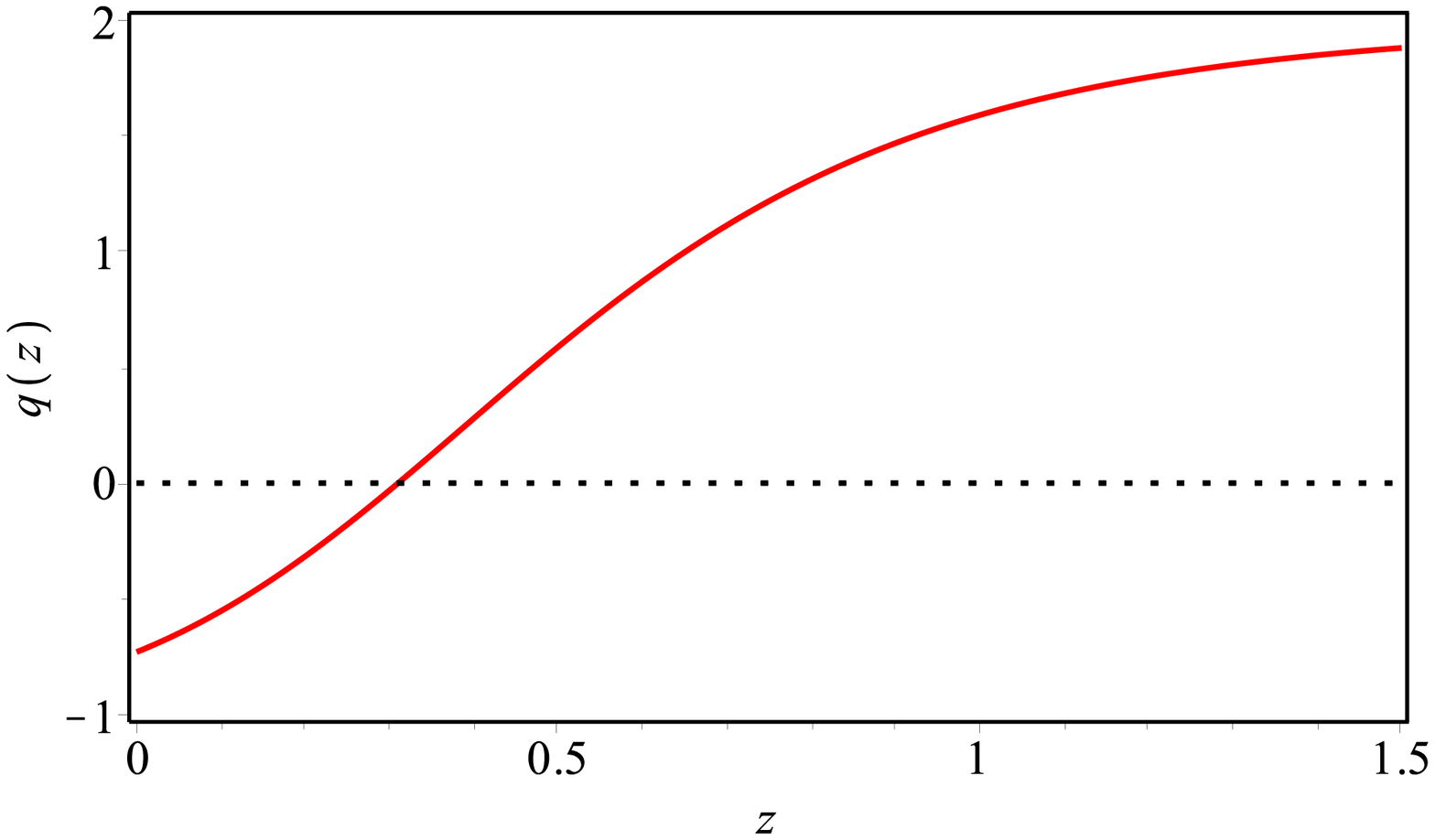}
\caption{The deceleration parameter of the external dimensions vs. cosmic redshift $z$. It is plotted by choosing $q_{a}=-0.73$ at $z=0$. The transition redshift to the accelerating expansion is $z_{{\rm t}}=0.31$.}
\label{fig:qz}
\end{minipage}
\end{figure}

Now we can talk about the world as seen by an observer living in four dimensions. The universe starts at $t=0$ from a singularity with $H_{a}=\infty$ and infinitely large energy density $\tilde{\rho}=\infty$ (or $\tilde{\rho}'=\infty$), that is, at the beginning there is a Big Bang. The universe then evolves from decelerating expansion to accelerating expansion, passing through different epochs where the effective fluid behaves differently; $a\sim t^{\frac{1}{3}}$ and $\tilde{w}\sim 1$ (stiff fluid dominated era) at very early times $t\sim 0$ and through a sequence of epochs where $a\sim t^{\frac{1}{2}}$ and $\tilde{w}\sim \frac{1}{3}$ (radiation dominated era), $a\sim t^{\frac{2}{3}}$ and $\tilde{w}\sim 0$ (pressureless matter dominated era), $a\sim t$ and $\tilde{w}\sim -\frac{1}{3}$ (acceleration starts at $t=\frac{1}{2\sqrt{\lambda}}\ln{(5+2\sqrt{6})}$) and reaches the present universe $a\sim t^{3.7}$ and $\tilde{w}\sim -0.82$. It eventually evolves to the de Sitter universe, $a\sim e^{\frac{\sqrt{\lambda}}{3}\,t}$ and $\tilde{w}\sim -1$, at the very late times (however, note that the universe is not effectively 4-dimensional at this epoch). One may form a judgement on the evolution sequence of the effective four dimensional universe from the behavior of the dimensionless deceleration parameter $q_{a}$. We depict the $q_{a}$ versus cosmic time $t$ in Fig. \ref{fig:dps} by using $\lambda=0.0187$, which gives the value $q_{a}=-0.73$ for the present-day universe. Such an evolution sequence is consistent with the current understanding of the universe, excluding the very far future of the universe.

As regards the present acceleration of the universe, the evolution
of the deceleration parameter with the cosmic redshift
$z=-1+\frac{a_{z=0}}{a}$ (where $a_{z=0}$ is the present value of
the scale factor) is also important to check if our model is
consistent with cosmological observations:
\begin{equation}
q_{a}(z)=-1+3\frac{(1+z)^{6}}{(1+z)^{6}+\frac{a_{z=0}^{6}}{a_{1}^{6}}}.
\end{equation}
We depict the deceleration parameter of the external dimensions
versus cosmic redshift $z$ by setting $q_{z=0}=-0.73$ in Fig.
\ref{fig:qz}. One may observe that $q_{a}=0$ at $z=z_{\rm t}=0.31$,
i.e., the accelerated expansion starts at $z_{\rm t}=0.31$, which is
in the range $0.3\lesssim z_{\rm t}\lesssim 0.8$ given in different observational studies \cite{Sahni03b,Rapetti07,GongWang07,CaiTuo11,Capozziello11,Capozziello97,Cunha09,Avgoustidis09,Komatsu11}.

As we are concerned with the recent transition from deceleration to
acceleration, it is also useful to take the third derivative of
scale factor of the observed universe into account. A convenient
parameter is the dimensionless jerk parameter $j$ that gives
opportunity to compare cosmological models with the $\Lambda$CDM
model in which it is constant $j_{\Lambda {\rm CDM}}=1$ \cite{Sahni03a,Sahni03b,Blandford04,Visser04,Rapetti07,Dunajski08}. In our model, on the other hand, the jerk parameter of the external space is dynamical:
\begin{equation}
j_{a}=\frac{\dddot{a}}{aH_{a}^3}=1+9{\rm
sech}^{2}(\sqrt{\lambda}\;t),
\end{equation}
which goes from $10$ to $1$ as the universe evolves. Using
$\lambda=0.0187$ we obtain for the present value of the jerk
parameter $j_{a}(13.7)=1.81$ which is also consistent with the
observational studies \cite{Rapetti07,Visser04}.

In short, using $\lambda=0.0187$, the internal dimensions are today
still at Planck length scales hence the observed universe is today
effectively four dimensional, it starts accelerating at $t_{\rm
t}=8.38$ (Gyr), i.e., acceleration starts $t_{0}-t_{\rm t}=5.32$
(Gyr) ago from now, the transition redshift is $z_{\rm t}=0.31$,
today $q_{a}=-0.73$ and $j_{a}=1.81$. Such a picture of the
universe is consistent with the observational studies.

\section{Final remarks}

It should be emphasized that our model doesn't involve a
cosmological constant $\Lambda$. The dynamical evolution of the external
(physical) and internal spaces are correlated and controlled by a
single real parameter $\lambda$ [see Eqn.(\ref{eqn:constraint})]. An
observer living in the 3-dimensional external space sees an
effective cosmic fluid with a specific time dependent EoS parameter
that drives the accelerated expansion of the universe and hence the
so-called cosmological constant problem doesn't arise here.

We also note that both the actual higher dimensional fluid and our
effective fluid in four dimensions involve time dependent EoS
parameters that start from $w= 1$ (stiff fluid) at very early times
and approach $w=-1$ (cosmological constant) at very late times. This
is exactly the type of behavior one would expect if a DE component
in four dimensional conventional general relativity without
cosmological constant had been introduced. A similar behavior is
obtained, for instance, for a quintessence field $\phi$ with a
constant potential $V(\phi)= \frac{\lambda}{3 \tilde{\kappa}_{0}}$ in
four dimensional conventional general relativity without
cosmological constant \cite{Demianski92}.

We also would like to note that our effective four dimensional model
induced from higher dimensions gives a more complete picture of our
current understanding of the universe compared with the standard
$\Lambda$CDM model. The $\Lambda$CDM model contains a binary
mixture of pressure-less matter (including CDM) and a positive cosmological
constant $\Lambda$. On the other hand, our four dimensional
effective universe exhibits a behavior expected of a four
dimensional universe in the presence of a certain mixture of stiff
matter, radiation, pressure-less matter (including CDM) and a cosmological
constant. A stiff fluid is the most promising EoS of matter at
ultra-high densities for representing the very early universe (see
\cite{Zeldovich62,Barrow78}). As the universe evolves, the matter
content becomes less stiff and the universe evolves into the
radiation dominated phase as should be expected.

As a final remark, we gave analytical solutions in $1+3+3$
dimensions. The number of internal dimensions may provide another
free parameter in the sense that more precise predictions (albeit
numerical) might be possible if we keep $ n \geq 3$ in our coupled
equations as a free parameter.

\begin{center}
\textbf{Acknowledgments}
\end{center}
\"{O}zg\"{u}r Akarsu and Tekin Dereli appreciate the financial support given by the Turkish Academy of Sciences (T{\"{U}}BA). \"{O}. Akarsu acknowledges also the financial support he is receiving from Ko\c{c} University.

\end{document}